\documentclass[iop,apj,tighten,twocolappendix,numberedappendix]{emulateapj}
\usepackage[breaklinks,colorlinks,citecolor=blue,linkcolor=red]{hyperref}

\usepackage{apjfonts}

\usepackage{graphicx}
\usepackage{amsmath}
\usepackage{amssymb}
\usepackage{color}
\usepackage{mathtools}
\usepackage{ulem}

\usepackage[all]{hypcap}

\newcommand\msun{\, {M}_\odot}

\newcommand\kms{\, {\rm km}\,{\rm s}^{-1}}

\newcommand\mseed{\mathcal{M}}
\newcommand\mcl{{M_{\rm CL}}}
\newcommand\ncl{{n_{\rm CL}}}
\newcommand\rhocl{{\rho_{\rm CL}}}
\newcommand\zcl{{Z_{\rm CL}}}
\newcommand\vesc{{v_{\rm esc}}}

\def\apjl{ApJL}%
\def\apjs{ApJS}%
\def\aap{A\&A}%
\begin{document}

\title{Repeated mergers, mass-gap black holes, and formation of intermediate-mass black holes in dense massive star clusters}

\author{Giacomo Fragione\altaffilmark{1,2}, Bence Kocsis \altaffilmark{3}, Frederic A. Rasio\altaffilmark{1,2}, Joseph Silk\altaffilmark{4,5,6}}
 \affil{$^1$Center for Interdisciplinary Exploration \& Research in Astrophysics (CIERA), Evanston, IL 60202, USA} 
  \affil{$^2$Department of Physics \& Astronomy, Northwestern University, Evanston, IL 60202, USA}
  \affil{$^3$Rudolf Peierls Centre for Theoretical Physics, Clarendon Laboratory, Parks Road, Oxford OX1 3PU, UK}
  \affil{$^4$Institut d’Astrophysique de Paris, UMR7095: CNRS \& UPMC, Sorbonne University, Paris, France}
   \affil{$^5$Department of Physics and Astronomy, The Johns Hopkins University, Baltimore, MD, USA}
   \affil{$^6$ BIPAC, Department of Physics, University of Oxford, Oxford, UK}

\begin{abstract}
Current theoretical models predict a mass gap with a dearth of stellar black holes (BHs) between roughly $50\,M_\odot$ and $100\,M_\odot$, while, above the range accessible through massive star evolution, intermediate-mass BHs (IMBHs) still remain elusive. Repeated mergers of binary BHs, detectable via gravitational wave emission with the current LIGO/Virgo/Kagra interferometers and future detectors such as LISA or the Einstein Telescope, can form both mass-gap BHs and IMBHs. Here we explore the possibility that mass-gap BHs and IMBHs are born as a result of successive BH mergers in dense star clusters. In particular, nuclear star clusters at the centers of galaxies have deep enough potential wells to retain most of the BH merger products after they receive significant recoil kicks due to anisotropic emission of gravitational radiation. Using for the first time simulations that include full stellar evolution, we show that a massive stellar BH seed can easily grow to $\sim 10^3 - 10^4\msun$ as a result of repeated mergers with other smaller BHs. We find that lowering the cluster metallicity leads to larger final BH masses. We also show that the growing BH spin tends to decrease in magnitude with the number of mergers, so that a negative correlation exists between final mass and spin of the resulting IMBHs. Assumptions about the birth spins of stellar BHs affect our results significantly, with low birth spins leading to the production of a larger population of massive BHs.
\end{abstract}

\section{Introduction}
\label{sect:intro}

The LIGO/Virgo/KAGRA (LVK) Collaboration has recently released the second Gravitational Wave Transient Catalog \citep[GWTC-2, ][]{lvc2020cat}, which, together with results from the first two observing runs \citep[GWTC-1, ][]{lvc2019cat}, is comprised of more than $50$ events. These detections are transforming our understanding of compact objects and gravitational wave (GW) physics \citep{lvc2020catb,lvc2020catc}. Over the coming years and decades, current (LIGO, Virgo, and KAGRA) and future (e.g., LISA, the Einstein Telescope, and DECIGO) GW detectors promise to provide unprecedented number of detections of black holes (BHs) and neutron stars (NSs).

The origin of binary mergers is still highly uncertain, with several possible scenarios that could potentially account for most of the observed events. These include mergers from isolated binary star evolution \citep[e.g.,][]{DominikBelczynski2013,bel16b,demi2016,ChruslinskaBelczynski2018,gm2018,SperaMapelli2019,Breivik2020,BaveraFragos2021,SantoliquidoMapelli2021,TanikawaKinugawa2021}, dynamical formation in dense star clusters \citep[e.g.,][]{PortegiesZwartMcMillan2000,askar17,baner18,fragk2018,KremerChatterjee2018,rod18,sams18,ham2019,krem2019,RastelloAmaro-Seoane2019,AntoniniGieles2020,DiCarloMapelli2020,FragioneBanerjee2020,RodriguezKremer2020,FragioneLoeb2021}, mergers in triple and quadruple systems \citep[e.g.,][]{antoper12,ll18,GrishinPerets2018,arcasedda+2018,Rodriguez2018a,fragg2019,flp2019,fragk2019,liu2019,FragioneLoeb2020,MartinezFragione2020,HamersFragione2021}, mergers of compact binaries in galactic nuclei \citep[e.g.,][]{oleary2009,bart17,PetrovichAntonini2017,sto17,HamersBar-Or2018,HoangNaoz2018,Arca-SeddaGualandris2018,FragioneLeighP2019,LiuLai20199,rasskoc2019,mck2020,Tagawa+2020,WangStephan2020,gondan21}. While several formation scenarios can account for some or even all of the merger rate, the contribution of different channels will hopefully be disentangled using a combination of the mass, spin, redshift, and eccentricity distributions as the number of detected events increases \citep[e.g.,][]{olea09,gondan2018,perna2019,ArcaSeddaMapelli2020,WongBreivik2020,MartinezRodriguez2021,SuLiu2021,zevin2021,TagawaKocsis2021b,TagawaKocsis2021a}.

One of the most interesting event in GWTC-2 is GW190521, a binary black hole (BBH) consistent with the merger of two BHs with masses of $91.4^{+29.3}_{-17.5} \msun$ and $66.9^{+15.5}_{-9.2} \msun$ \citep{ligo2020new1,ligo2020new2}. Stellar evolutionary models predict no BHs with masses larger than about $50 - 70\msun$ (high-mass gap), resulting from the pulsational pair-instability process which affects massive progenitors. Whenever the pre-explosion stellar core is in the range $45 - 65\msun$, large amounts of mass can be ejected, leaving a BH remnant with a maximum mass around $50 - 70\msun$ \citep{heger2003,woosley2017}. Recent studies have shown that the exact lower boundary of the high-mass gap depends on the stellar metallicity \citep{woosley2017,limongi2018,bel2020,VinkHiggins2021}; the upper boundary is around $125\msun$ whenever the metallicity is $\lesssim 10^{-3}$ \citep{spera2017,renzo2020}.

\begin{table*}
\caption{Description of important quantities used in the text.}
\centering
\begin{tabular}{lc}
\hline
Symbol & Description\\
\hline\hline
$\mcl$              & Cluster mass \\
$\rhocl$            & Cluster mass density \\
$\rho_{\rm BH}$            & BH mass density \\
$\ncl$            & Cluster number density \\
$n_{\rm BH}$            & BH number density \\
$\zcl$              & Cluster metallicity \\
$\vesc$             & Cluster escape speed \\
$\sigma$            & Cluster velocity dispersion \\
$\sigma_{\rm BH}$            & BH velocity dispersion \\
$f_{\rm BH}$    & Fraction of the cluster mass in BHs \\
\hline
$\mseed$            & Mass of the growing BH  \\
$\mseed_*$          & Mass of the stellar progenitor of the BH seed \\
$\mseed_{\rm ini}$  & Mass of the BH seed \\
$\beta_*$           & Slope of stellar initial mass function for $m_*>0.5\msun$ \\
$m_{\rm min, *}$       & Minimum mass of the stellar mass function \\
$m_{\rm max, *}$       & Maximum mass of the stellar mass function \\
$f_{\rm b, *}$         & Primordial binary fraction for massive stars ($m_*>20\msun$) \\
$q_*$                 & Mass ratio of primordial stellar binary \\
$a_*$                 & Semimajor axis of primordial stellar binary \\
$a_{\rm min, *}$    & Minimum semimajor axis of primordial stellar binary \\
$a_{\rm max, *}$    & Maximum semimajor axis of primordial stellar binary \\
$a_{\rm h, *}$        & Hard--soft boundary for primordial stellar binary \\
$e_*$         & Eccentricity of primordial stellar binary \\
$v_{\rm natal}$     & Natal kick imparted at BH formation \\
$\nu$               & Velocity dispersion of the Maxwellian distribution of natal kicks \\
$\chi$              & BH dimensionless spin parameter\\
$\chi_\mathcal{M}$          & Spin parameter of BH seed\\
\hline
$m_{\rm min, BH}$   & Minimum mass of the BH mass function \\
$m_{\rm max, BH}$   & Maximum mass of the BH mass function \\
$\beta_{\rm BH}$    & Slope of the BH mass function \\
$a_{\rm ej}$   & Semimajor axis of the BH binary corresponding to ejection from the parent cluster \\
$a_{\rm GW}$   & Semimajor axis of the BH binary when GW emission takes over \\
$e_{\rm BH}$        & Eccentricity of the BH binary \\
\hline
$\tau_{\rm DF, *}$     & Dynamical friction timescale for the BH seed progenitor \\
$\tau_{\rm DF}$     & Dynamical friction timescale for the BH seed \\
$\tau_{\rm 3bb}$     & Timescale for binary formation via three-body interaction \\
$\tau_{\rm 1}^*$     & Timescale for binary formation via binary--single interaction mediated by stellar binaries \\
$\tau_{\rm 1}$     & Timescale for binary formation via binary--single interaction \\
$\tau_{\rm 2}$     & Timescale to shrink the BH binary to $\max(a_{\rm ej},a_{\rm GW})$ \\
$\tau_{\rm GW}$    & Timescale to merge via GW emission \\
\hline
$v_{\rm kick}$      & Recoil kick due to anisotropic GW emission  \\
$\mathcal{M}_{\rm fin}$      & Mass of the merger remnant \\
$\chi_{\rm fin}$      & Spin parameter of the merger remnant \\
\hline
\end{tabular}
\label{tab:quant}
\end{table*}

BHs more massive than the limit imposed by the pulsational pair instability can be produced dynamically in the core of a dense star cluster. Here, three- and four-body interactions can catalyze the growth of a BH seed through repeated mergers with smaller BHs \citep[e.g.,][]{gultek2004,antonini2019,BaibhavGerosa2020,frsilk2020,MapelliDall'Amico2021}. A fundamental limit for hierarchical mergers comes from the recoil kick imparted to merger remnants \citep[e.g.][]{lou10,lou11,VarmaGerosa2019}. Depending on the mass ratio and the spins of the merging BHs, the recoil kick can be as high as $\sim 100 - 1000\kms$, and could results in the ejection of the merger remnant if it exceeds the local escape speed. However, if the host cluster hosts a supermassive black hole (SMBH), or if the cluster does not have an SMBH but it is massive and dense enough, as in nuclear star clusters (NSCs) or the most massive globular clusters (GCs), hierarchical mergers can build up BHs in the mass gap and even form intermediate-mass black holes (IMBHs), possibly explaining the formation of GW19052-like events \citep[e.g.,][]{AntoniniGieles2020,FragioneLoeb2020,frsilk2020,BaibhavBerti2021,MapelliDall'Amico2021,TagawaKocsis2021a,TagawaKocsis2021c}.

While interesting, repeated mergers are computationally expensive to investigate in detail and over broad ranges of initial conditions with full $N$-body simulations \citep[e.g.,][]{Aarseth2003,gier2006MNRAS.371..484G,Pattabiraman2013}. Therefore, developing an alternative and more rapid method is highly desirable. In this paper, we present a semi-analytic framework to investigate hierarchical mergers in dense star clusters, which expands upon the method originally developed in \citet{frsilk2020}. Our approach allows us to rapidly probe how the outcomes of hierarchical mergers in dense star clusters are affected by cluster masses, stellar densities, and metallicities, and by different assumptions on the BH mass spectrum and spins. Importantly, we have also integrated for the first time the updated versions of the stellar evolution codes \textsc{sse} and \textsc{bse} \citep{HurleyPols2000,HurleyTout2002}, including the most up-to-date prescriptions for stellar winds and compact object formation \citep{BanerjeeBelczynski2020}. This provides more realistic BH mass spectra and spins, and allows us to study the role of primordial binaries.

This paper is organized as follows. In Section~\ref{sect:method}, we discuss our numerical method to study repeated mergers, the formation of IMBHs, and BH mergers in the mass gap. In Section~\ref{sect:results}, we discuss the assembly of massive BHs through repeated mergers and study the role of the cluster metallicity, primordial binary fraction, and prescriptions for the remnant spin. Finally, in Section~\ref{sect:concl}, we discuss the implications of our results and draw our conclusions.

\section{Method}
\label{sect:method}

In what follows, we describe the details of the numerical method we use to follow the evolution of a BH of seed mass $\mseed$, which undergoes mergers with other BHs.

\subsection{Cluster properties}

The characteristics of the host NSC are essentially determined by its mass $\mcl$, its half-mass radius $r_{\rm h}$, and core radius $r_{\rm c}$. We assume that the cluster density is described by a three-parameter potential-density pair described in \citet{StoneOstriker2015}
\begin{equation}
\rho_{\rm CL}=\frac{\rho_{\rm c}}{(1+r^2/r_{\rm c}^2)(1+r^2/r_{\rm h}^2)}\,,
\end{equation}
where
\begin{equation}
\rho_{\rm c}=\frac{M_{\rm CL}(r_{\rm h}+r_{\rm c})}{2\pi^2r_{\rm c}^2r_{\rm h}^2}
\end{equation}
is the central density, $r_{\rm c}$ is the core radius, and $r_{\rm h}$ is the half-mass density radius when $r_{\rm h}\gg r_{\rm c}$. these models are designed as an analytically tractable alternative to single-mass King models. The escape velocity from the center is\footnote{We assume an average compactness parameter $r_{\rm h}/r_{\rm c}=10$ \citep{GeorgievBoker2014}.}
\begin{eqnarray}
\vesc&=& 2\sqrt{\frac{\log(r_{\rm h}/r_{\rm c})}{\pi}}\left(\frac{GM_{\rm CL}}{r_{\rm h}- r_{\rm c}}\right)^{1/2} \nonumber\\
& \approx & 50\kms\left(\frac{M_{\rm CL}}{10^5\msun}\right)^{1/2}\left(\frac{r_h}{1\,{\rm pc}}\right)^{-1/2}\,,
\label{eqn:mvesc}
\end{eqnarray}
and the core velocity dispersion is computed as
\begin{equation}
\sigma=\frac{3(\pi^2-8)}{8}\vesc\,.
\end{equation}
Finally, the cluster metallicity, $\zcl$, determines the BH mass spectrum. The velocity dispersion and densities of BHs in the core can be related to the corresponding quantities for the background stars. Stellar and BH velocity dispersions are related by
\begin{equation}
\xi^{-2}=\frac{\langle m_{\rm BH}\rangle\sigma_{\rm BH}^2}{\langle m_{\rm *}\rangle \sigma^2}\,,
\end{equation}
where $\langle m_{\rm *}\rangle\approx 0.5\msun$ and $\langle m_{\rm BH}\rangle\approx 10\msun$, which quantifies the deviation from energy equipartition between stars and BHs. We fix $\xi=1/\sqrt{5}$, which gives timescales for BBH formation consistent with the numerical study of \citet{mors2015}. When the BH population dominates the mass in the core, its density is \citep{lee1995,ChoksiVolonteri2019}
\begin{equation}
\rho_{\rm BH}=\frac{1}{2}\left(\frac{r_{\rm c}}{r_{\rm h}}\right)^2 f_{\rm BH}^{-2}\,\rho_{\rm c}\approx\frac{f_{\rm BH}^{-2}}{4\pi^2}\frac{M_{\rm CL}}{r_{\rm h}^3}\,,
\end{equation}
where $f_{\rm BH}=0.01$ is the fraction of the cluster mass in BHs.

Note that we assume that the masses and sizes of NSCs do not evolve significantly during their lifetime. Observations show that NSCs in general tend to have a wide range of stellar ages, including young stellar populations \citep[e.g.,][]{RossavanderMarel2006,SethDalcanton2006}. Indeed, star clusters lose mass, expand as a consequence of two-body relaxation, are continuously supplied with stars and gas from the rest of the galaxy, and accrete star clusters. For a more  detailed discussion of the possible effects of NSC evolution, see Sect.~\ref{sect:concl}.

\subsection{Population of nuclear star clusters}

To generate a population of NSCs, we start from sampling galaxy masses from a Schechter function
\begin{equation}
    \Phi(M_{\rm *,gal}) \propto \left(\frac{M_{\rm *,gal}}{M_{\rm c}} \right)^{\alpha_{\rm c}} \exp\left(-\frac{M_{\rm *,gal}}{M_{\rm c}}\right)\,,
\end{equation}
where $M_{\rm *,gal}$ is the stellar mass of a given galaxy. We set $M_{\rm c} = 10^{11.14}\msun$ and $\alpha_{\rm c}=1.43$, as extracted from the EAGLE cosmological simulations in \citet{FurlongBower2015}. We then use scaling relations for late-type galaxies from \citet{georg2016} to scale the galaxy mass to the NSC mass, 
\begin{equation}
    \log(M_{\rm CL}/c_1)=\zeta\times\log(M_{\rm *, gal}/c_2)+\psi
    \label{eqn:mclmgal},
\end{equation}
and to draw the half-mass radius
\begin{equation}
    \log(r_{\rm h}/c_3)=\kappa\times\log(M_{\rm CL}/c_4)+\omega\,,
    \label{eqn:rhmcl}
\end{equation}
where $c_1= 2.78\times 10 ^6\msun$, $c_2= 3.94\times 10^9\msun$, $\zeta = 1.001^{+0.054}_{-0.067}$, $\psi=0.016^{+0.023}_{-0.061}$ and $c_3=3.31$~pc, $c_4=3.60\times 10^6\msun$, $\kappa=0.321^{+0.047}_{-0.038}$, $\omega=-0.011^{+0.014}_{-0.031}$. In sampling from Eqs.~\ref{eqn:mclmgal}-\ref{eqn:rhmcl}, we consider the scatter in the fit parameters.

\subsection{Black hole seeds}

To compute the initial BH seed mass $\mseed_{\rm ini}$, we first sample the mass of its stellar progenitor, $\mseed_*$, from the canonical initial mass function \citep{kro01},
\begin{equation}
\xi(m_*)=k_1
\begin{cases}
\left(\frac{m_*}{0.5}\right)^{-1.3}& \text{$0.08\le m_*/\mathrm{M}_\odot\leq 0.50$},\\
\left(\frac{m_*}{0.5}\right)^{-\beta_*}& \text{$0.50\le m_*/\mathrm{M}_\odot\leq 150.0$},
\end{cases}
\label{eqn:imf}
\end{equation}
where $k_1\approx 0.62$ is a normalization factors for $\beta_*=2.35$. The mass of the BH seed progenitor is drawn in the range $[m_{\rm min, *}, m_{\rm max, *}]$, where $m_{\rm min, *}=20\,\msun$ and $m_{\rm max, *}=150\,\msun$, respectively, which roughly encompass the possible masses of BH progenitors. We then evolve the progenitor mass $\mseed_*$ at a metallicity $\zcl$ using the stellar evolution code \textsc{sse} \citep{HurleyPols2000}. Our current version of \textsc{sse} includes the most up-to-date prescriptions for stellar winds and remnant formation \citep[see][and references therein]{BanerjeeBelczynski2020}. Alternatively, the initial BH seed mass $\mseed_{\rm ini}$ can be fixed to a specific value. One possibility is that the initial BH seed is the result of a runaway growth and its mass is typically $\lesssim 1\%$ of the cluster mass \citep{por02}.

If the primordial binary fraction in high-mass stars, $f_{\rm b_*}$, is sufficiently high, the stellar progenitor, $\mseed_*$, can be born in a binary system. In this case, we sample the mass of the companion using a uniform distribution in the mass ratio $q_*$, consistent with observations of massive binary stars \citep{SanadeMink2012,DucheneKraus2013,Sana2017}
\begin{equation}
    f(q_*)\propto \rm const\,.
\label{eqn:mratio}
\end{equation}
We draw the semi-major axis $a_*$ using a log-uniform distribution, roughly consistent with the observational study of \citet{KobulnickyKiminki2014},
\begin{equation}
    f(a_*)\propto \frac{1}{a_*}\,
\end{equation}
in the range $[a_{\rm min}, a_{\rm max}]$. We fix the minimum semi-major axis to $0.1$ AU, while the maximum semi-major axis is set to be the hard-soft boundary, defined as
\begin{equation}
    a_{h}=800\ \mathrm{AU}\left(\frac{q_*\mseed_*^2}{400\,{\rm M}_\odot^2}\right)\left(\frac{\sigma}{30\kms}\right)^{-2}\,.
\label{eqn:ah}
\end{equation}
Finally, we sample the binary eccentricity, $e_*$, from a thermal distribution \citep{Heggie1975}.

The primordial binary is then evolved using \textsc{bse} \citep{HurleyTout2002}, which includes the most up-to-date prescriptions for stellar winds and remnant formation \citep[see][and references therein]{BanerjeeBelczynski2020}. There are three possible outcomes of binary stellar evolution: i) the binary components merge; ii) a binary BH is formed; iii) the binary disrupts as a result of stellar evolution and/or natal kicks, which we assume are imparted to the compact object remnant when the progenitor star collapses due to mass loss and supernova explosion. For simplicity, we only consider primordial binaries that lead to the formation of a binary BH. Note, however,  that recent results have shown that subsequent mergers catalyzed by massive binaries can produce massive BHs in the early phases of the cluster evolution \citep{KremerSpera2020,GonzalezKremer2021}. At the moment, we do not model this scenario, which we leave to a future study. We take the natal kick to follow a Maxwellian distribution
\begin{equation}
    p(v_{\rm natal}) \propto v_{\rm natal}^2\,e^{-v_{\rm natal}^2/\nu^2}
\label{eqn:vnat}
\end{equation}
with velocity dispersion $\nu = 265\kms$, based on observations of radio pulsars \citep{hobbs2005}. BH natal kicks are assigned assuming momentum conservation \citep{fryerkalo2001}; thus, the natal velocity of a BH of mass $m_{\rm BH}$ is lowered by a factor of $1.4\,\msun/ {m_{\rm BH}}$, with $1.4\,\msun$ being the typical neutron star mass. Besides unbinding the binary, natal kicks can eject the BH seed, either as a single or in a binary, from the parent cluster whenever they exceed the cluster escape speed. For what concerns the BH natal spins $\chi$, we consider two different models, where the prescriptions of the \textsc{geneva} stellar evolution code \citep{EggenbergerMeynet2008,EkstromGeorgy2012} and \textsc{mesa} stellar evolution code \citep{PaxtonBildsten2011,PaxtonMarchant2015} are used, respectively. Alternatively, the initial spin of BHs is assumed to vanish, consistent with the recent findings of \citet{FullerMa2019}.

\subsection{Dynamical interactions and mergers}
\label{subsect:dynamics}

Either the progenitor is born as a single or in a binary star, it sinks to the cluster center from its initial position via dynamical friction over a timescale \citep{Chandrasekhar1943}
\begin{equation}
\tau_{\rm df}\approx 17\ \mathrm{Myr} \left(\frac{20\msun}{(1+q_*)\mseed_*}\right)\left(\frac{\mathcal{R}}{1\,\mathrm{pc}}\right)^{3/2} \left(\frac{\mcl}{10^5\msun}\right)^{1/2}\,,
\label{eqn:tdf}
\end{equation}
where $\mathcal{R}$ is the distance from the cluster center. In the previous equation, $q_*=0$ in the case the seed progenitor is a single. As a typical initial distance, we choose $r_{\rm h}$, such that the initial dynamical friction timescale is $\tau_{\rm df}(r_{\rm h})$. As discussed, compact objects are imparted a natal kick that could eject them from the core of the parent cluster. Therefore, we check that the natal kicks imparted to the system are below the cluster escape speed, $v_{\rm esc}$. If not ejected from the cluster, the system sinks back to the cluster center over a dynamical friction timescale.

\begin{table*}
\caption{Model parameters: model number, star cluster, metallicity ($Z$), primordial binary fraction of massive stars ($f_{\rm b,*}$), spin model, initial seed mass ($\mseed_{\rm ini}$), fraction of models that produce a BH more massive than $100\,\msun$, fraction of models that produce a BH more massive than $500\,\msun$, fraction of models that produce a BH more massive than $1000\,\msun$.}
\centering
\begin{tabular}{ccccccccc}
\hline\hline
Model & Star cluster & $Z_{\rm CL}$ & $f_{\rm b,*}$ & Spin model & $\mseed_{\rm ini}$ ($\msun$) & $f(\mseed > 100\,\msun)$ & $f(\mseed > 500\,\msun)$ & $f(\mseed > 800\,\msun)$\\
\hline
1 & NSC & 0.0001 & 0.0 & \citet{FullerMa2019} & Stell. Evol. & $1.4\times 10^{-1}$ & $9.2\times 10^{-3}$ & $2.9\times 10^{-3}$ \\
1E & NSC & 0.0001 & 0.0 & \citet{FullerMa2019} & Stell. Evol. & $8.9\times 10^{-2}$ & $2.3\times 10^{-3}$ & $1.1\times 10^{-3}$ \\
2 & NSC & 0.001 & 0.0 & \citet{FullerMa2019} & Stell. Evol. & $8.7\times 10^{-2}$ & $6.7\times 10^{-3}$ & $2.0\times 10^{-3}$ \\
3 & NSC & 0.01 & 0.0 & \citet{FullerMa2019} & Stell. Evol. & $1.1\times 10^{-2}$ & $2.2\times 10^{-4}$ & $0$ \\
4 & NSC & 0.0001 & 0.25 & \citet{FullerMa2019} & Stell. Evol. & $1.4\times 10^{-1}$ & $9.9\times 10^{-3}$ & $4.8\times 10^{-3}$ \\
5 & NSC & 0.0001 & 0.5 & \citet{FullerMa2019} & Stell. Evol. & $1.3\times 10^{-1}$ & $9.1\times 10^{-3}$ & $4.2\times 10^{-3}$ \\
6 & NSC & 0.0001 & 0.75 & \citet{FullerMa2019} & Stell. Evol. & $1.3\times 10^{-1}$ & $9.5\times 10^{-3}$ & $3.9\times 10^{-3}$ \\
7 & NSC & 0.0001 & 1.0 & \citet{FullerMa2019} & Stell. Evol. & $1.3\times 10^{-1}$ & $7.6\times 10^{-3}$ & $3.1\times 10^{-3}$ \\
8 & NSC & 0.0001 & 0.0 & \textsc{geneva} & Stell. Evol. & $9.2\times 10^{-3}$ & $0$ & $0$ \\
9 & NSC & 0.0001 & 0.0 & \textsc{mesa} & Stell. Evol. & $9.6\times 10^{-2}$ & $6.9\times 10^{-3}$ & $2.2\times 10^{-3}$ \\
10 & NSC & 0.0001 & 0.0 & \citet{FullerMa2019} & $50$ & $5.6\times 10^{-1}$ & $4.3\times 10^{-2}$ & $2.1\times 10^{-2}$ \\
11 & NSC & 0.0001 & 0.0 & \citet{FullerMa2019} & $100$ & $1$ & $1.6\times 10^{-1}$ & $1.1\times 10^{-1}$ \\
12 & NSC & 0.0001 & 0.0 & \citet{FullerMa2019} & $150$ & $1$ & $2.8\times 10^{-1}$ & $2.2\times 10^{-1}$ \\
13 & NSC & 0.0001 & 0.0 & \citet{FullerMa2019} & $200$ & $1$ & $4.1\times 10^{-1}$ & $3.4\times 10^{-1}$ \\
14 & GC & 0.0001 & 0.0 & \citet{FullerMa2019} & Stell. Evol. & $3.5\times 10^{-2}$ & $0$ & $0$ \\
15 & GC & 0.001 & 0.0 & \citet{FullerMa2019} & Stell. Evol. & $1.5\times 10^{-2}$ & $0$ & $0$ \\
16 & GC & 0.01 & 0.0 & \citet{FullerMa2019} & Stell. Evol. & $0$ & $0$ & $0$ \\
\hline\hline
\end{tabular}
\label{tab:models}
\end{table*}

In the case that the BH seed was born with a companion, stellar evolution and natal kicks can unbind the binary, leaving the BH seed as a single. In the case that binary remains bound, the stellar evolutionary processes could produce a binary BH, whose semi-major axis exceeds the hard-soft boundary (see Eq.~\ref{eqn:ah}). Any interaction with a third BH or star will tend to make the binary even softer and, eventually, disrupt it, leaving two single BHs \citep{Heggie1975}. In these case, where we simply keep following the further evolution of the most massive BH, or in the case the BH seed was born as a single star, a new BH companion can be found in the core of the dense cluster via dynamical friction. The mass of $m_2$ is drawn assuming that the pairing probability scales as $\mathcal{P}\propto (\mseed+m_2)^{\beta_{\rm BH}}$. We set $\beta_{\rm BH}=4$, based on numerical simulations of globular clusters \citep{omk16}, and sample the secondary mass in the range $[m_{\rm min, BH}, m_{\rm max, BH}]$, as appropriate for the BH mass spectrum at metallicity $Z_{\rm CL}$.

Interactions to form a binary BH come in two flavors: encounters between three single objects and encounters between a single and a binary. The typical timescale for the former is \citep[e.g.,][]{lee1995}
\begin{eqnarray}
\tau_{\rm 3bb}&=&125\ \mathrm{Myr}\left(\frac{10^6\ \mathrm{pc}^{-3}}{n_{\rm BH}}\right)^2\times \nonumber\\
&\times& \left(\xi^{-1} \frac{\sigma}{30\kms}\right)^9 \left(\frac{20\msun}{\mseed}\right)^5\,,
\label{eqn:t3bb}
\end{eqnarray}
where $n_{\rm BH}= \rho_{\rm BH}/\langle m_{\rm BH}\rangle$ is the number density of BHs near the center. Formation of BBHs via exchange binary-single encounters are first mediated by stellar binaries, which, assuming an overall stellar binary fraction of $0.05$, happen over a timescale \citep{antoras2016}
\begin{eqnarray}
\tau_1^*&=&220\ \mathrm{Myr}\ \left(\frac{10^6\ \mathrm{pc}^{-3}}{n_{\rm c}}\right)\left(\frac{\sigma}{30\kms}\right)\times\nonumber\\
&&\times \left(\frac{20\msun}{\mseed+2m_*}\right)\left(\frac{1\ \mathrm{AU}}{a_h^*}\right)\,,
\label{eqn:t11}
\end{eqnarray}
where $n_{\rm c}=\rho_{\rm c}/\langle m_{\rm *}\rangle$ is the stellar number density near the center and
\begin{equation}
a_h^*=2\ \mathrm{AU}\left(\frac{m_*}{1\msun}\right)^2\left(\frac{\sigma}{30\kms}\right)^{-2}\,.
\end{equation}
is the hard-soft semi-major axis for a stellar binary. In our calculation, we assume $m_*=0.5\,\msun$. Once BH binaries are formed, they dominate the dynamics inside the cluster core and, assuming a BH binary fraction of $0.01$ \citep{mors2015}, binary--single encounters between a single BH and a BH binary occur on a timescale \citep{mill2009,antoras2016}
\begin{eqnarray}
\tau_1&=&450\ \mathrm{Myr}\ \xi^{-1}\left(\frac{10^6\ \mathrm{pc}^{-3}}{n_{\rm BH}}\right)\left(\frac{\sigma}{30\kms}\right)\times\nonumber\\
&\times& \left(\frac{20\msun}{m_3+\mseed+m_2}\right)\left(\frac{1\ \mathrm{AU}}{a_h}\right)\,,
\label{eqn:t12}
\end{eqnarray}
where $m_3$ is the mass of the third BH taking part in the interaction, which we set to $10\,\msun$ in our calculations.

We then compute the further evolution of a binary BH that is dynamically assembled or that originates from a primordial stellar binary (not unbound by stellar evolution or natal kicks). We assume that the BH binary shrinks at a constant rate (see Eq.~\ref{eqn:t2}), eventually to the regime where GWs take over \citep{quin1996}. During any of the interactions, the binary could receive a recoil kick high enough to be ejected, which happens when \citep{antoras2016}
\begin{equation}
a_{\rm ej}>a_{\rm GW}\,,
\label{eqn:ejok}
\end{equation}
where
\begin{eqnarray}
a_{\rm ej}&=&0.2\ \mathrm{AU}\left(\frac{400\ \mathrm{M}_\odot^2}{(\mseed+m_2)(m_3+\mseed+m_2)}\right)\times\nonumber\\
&\times&\left(\frac{\mu}{10\msun}\right)\left(\frac{50\kms}{\vesc}\right)^2\,,
\label{eqn:aej}
\end{eqnarray}
with $\mu$ being the reduced mass of the $\mseed-m_2$ system, and
\begin{eqnarray}
a_{\rm GW}&=&0.08\ \mathrm{AU}\left(\frac{\mseed+m_2}{20\msun}\right)^{3/5}\left(\frac{10^5\msun\ \mathrm{pc}^{-3}}{\rho_{\rm BH}}\right)^{1/5}\times \nonumber\\
&\times& \left(\frac{\sigma}{30\kms}\right)^{1/5}\left(\frac{q}{(1+q)^2}\right)^{1/5}\,,
\label{eqn:agw}
\end{eqnarray}
with $q=m_2/\mseed$. Assuming that each interaction removes a fraction $0.2 m_3/(\mseed+m_2)$ of the binary binding energy \citep{quin1996}, the binary will shrink until reaching $\max(a_{\rm ej},a_{\rm GW})$ over a timescale \citep{millh2002}
\begin{equation}
\tau_2\approx 5\left(\frac{m_3}{\mseed + m_2}\right)^{-1}\tau_{21}\,,
\label{eqn:t2}
\end{equation}
where
\begin{eqnarray}
\tau_{21}&=&20\ \mathrm{Myr}\ \xi^{-1}\left(\frac{10^6\ \mathrm{pc}^{-3}}{n_{\rm BH}}\right)\left(\frac{\sigma}{30\kms}\right)\times\nonumber\\
&\times &\left(\frac{0.05\ \mathrm{AU}}{\max(a_{\rm ej},a_{\rm GW})}\right)\left(\frac{20\msun}{\mseed+m_2}\right)\,.
\label{eqn:t2}
\end{eqnarray}
We sample all the relevant timescales from a Poisson distribution, that is $\exp(-t/\Theta)$. In particular, $\Theta=\min(\tau_{\rm 3bb}, \tau_1)$ is the mean time after which the seed forms a BBH, while $\Theta=\tau_2$ is the mean time after which the BBH shrinks to $\max(a_{\rm ej},a_{\rm GW})$. Finally, the binary merges over a timescale \citep{peters64}
\begin{eqnarray}
\tau_{\rm GW}&=&250\ \mathrm{Myr}\left(\frac{8000\msun}{\mseed m_2(\mseed+m_2)}\right)\times \nonumber\\
&\times& \left(\frac{\max(a_{\rm ej},a_{\rm GW})}{0.05\ \mathrm{AU}}\right)^4 (1-e_{\rm BH}^2)^{7/2}\,,
\label{eqn:tgw}
\end{eqnarray}
where $e_{\rm BH}$ is the eccentricity, which we sample from a thermal distribution \citep{jeans1919,Heggie1975}. If $a_{\rm ej}>a_{\rm GW}$, the binary is ejected from the cluster halting further growth, otherwise the merger happens in cluster.

\subsection{Recoil kicks and merger remnants}

The merger remnant receives a recoil kick as a result of the anisotropic emission of GWs at merger and can be ejected from the host star cluster. This GW recoil kick depends on the asymmetric mass ratio $\eta=q/(1+q)^2$ and on the magnitude of the dimensionless spin parameters, $|\boldsymbol{{\chi_1}}|=|\boldsymbol{{\chi_\mseed}}|$ and $|\boldsymbol{{\chi_2}}|$. In our models, spin orientations are assumed to be isotropic, as appropriate for merging binaries assembled dynamically. We model the recoil kick as \citep{lou10,lou12}
\begin{equation}
\textbf{v}_{\mathrm{kick}}=v_m \hat{\textbf{e}}_{\perp,1}+v_{\perp}(\cos \xi \hat{\textbf{e}}_{\perp,1}+\sin \xi \hat{\textbf{e}}_{\perp,2})+v_{\parallel} \hat{\textbf{e}}_{\parallel}\,,
\label{eqn:vkick}
\end{equation}
where
\begin{eqnarray}
v_m&=&A\eta^2\sqrt{1-4\eta}(1+B\eta)\\
v_{\perp}&=&\frac{H\eta^2}{1+q}(\chi_{2,\parallel}-q\chi_{1,\parallel})\\
v_{\parallel}&=&\frac{16\eta^2}{1+q}[V_{1,1}+V_A \tilde{S}_{\parallel}+V_B \tilde{S}^2_{\parallel}+V_C \tilde{S}_{\parallel}^3]\times \nonumber\\
&\times & |\mathbf{\chi}_{2,\perp}-q\mathbf{\chi}_{1,\perp}| \cos(\phi_{\Delta}-\phi_{1})\,.
\end{eqnarray}
The $\perp$ and $\parallel$ refer to the direction perpendicular and parallel to the orbital angular momentum, respectively, while $\hat{e}_{\parallel, 1}$ and $\hat{e}_{\parallel, 2}$ are orthogonal unit vectors in the orbital plane. We have also defined the vector
\begin{equation}
\tilde{\mathbf{S}}=2\frac{\boldsymbol{\chi}_{2}+q^2\boldsymbol{\chi}_{1}}{(1+q)^2}\,,
\end{equation}
$\phi_{1}$ as the phase angle of the binary, and $\phi_{\Delta}$ as the angle between the in-plane component of the vector
\begin{equation}
\boldsymbol{\Delta}=M^2\frac{\boldsymbol{\chi}_{2}-q\boldsymbol{\chi}_{1}}{1+q}
\end{equation}
and the infall direction at merger. Finally, we adopt $A=1.2\times 10^4$ km s$^{-1}$, $H=6.9\times 10^3$ km s$^{-1}$, $B=-0.93$, $\xi=145^{\circ}$ \citep{gon07,lou08}, and $V_{1,1}=3678$ km s$^{-1}$, $V_A=2481$ km s$^{-1}$, $V_B=1793$ km s$^{-1}$, $V_C=1507$ km s$^{-1}$ \citep{lou12}. We adjust the final total spin of the merger product, $\chi_{\rm fin}$, and its mass, $\mathcal{M}_{\rm fin}$,  using the results of \citet{Jimenez-FortezaKeitel2017}, which we generalized to precessing spins using an approach similar to \citet{HofmannBarausse2016}. 

Whenever $v_{\rm kick}>v_{\rm esc}$, the remnant is ejected from the host cluster, while, if $v_{\rm kick}<v_{\rm esc}$, $\mseed$ will be retained. After a dynamical friction timescale (see Eq.~\ref{eqn:tdf}), the merged remnant sinks back to the center, eventually forming new binaries and merging, further growing in mass.

\subsection{Overview of the method and growth of a black hole seed}

We summarize the various steps of our method, which allow us to study the evolution and growth of a black hole seed. 

\begin{enumerate}
    \item We sample the stellar mass of the seed progenitor, $\mseed_*$, from Eq.~\ref{eqn:imf}. To decide whether the seed progenitor was born in a binary system, we extract a random number. If it is smaller than $f_{b,*}$, we sample the companion mass (Eq.~\ref{eqn:mratio}) and the orbital properties of the binary. Finally, we compute the timescale for the seed to sink to the cluster center (Eq.~\ref{eqn:tdf}). 
    
    \item We evolve the seed progenitor using \textsc{sse}, if it is a single, or using \textsc{bse}, if it is in a binary. If the the seed is not ejected by natal kicks (Eq.~\ref{eqn:vnat}), we compute the timescale for the seed to sink back to the cluster center (Eq.~\ref{eqn:tdf}).
    
    \item In the case that the BH seed, $\mseed$, is a single or the primordial binary and is disrupted as a result of stellar evolution leaving a single BH, we compute the timescale (Eqs.~\ref{eqn:t3bb}-\ref{eqn:t11}-\ref{eqn:t12}) to find a BH companion. The mass of the companion, $m_2$, is drawn assuming that the pairing probability scales as $\mathcal{P}\propto (\mseed+m_2)^4$.
    
    \item After the BBH is formed, or in the case the BBH is the result of binary stellar evolution of the primordial binary, we compute the timescale (Eq.~\ref{eqn:t2}) for it to shrink to $\max(a_{\rm ej},a_{\rm GW})$.
    
    \item If $a_{\rm ej}>a_{\rm GW}$, the BBH is ejected from the cluster, halting further growth. If $a_{\rm ej}<a_{\rm GW}$, we compute the timescale to merge via GW emission (Eq.~\ref{eqn:tgw}).
    
    \item We compute the recoil kick (Eq.~\ref{eqn:vkick}) and the remnant spin and mass. If it is not ejected, we compute the timescale for the seed to sink back to the cluster center (Eq.~\ref{eqn:tdf}).
    
    \item If the seed is retained in the cluster and the total elapsed time is smaller than $10$ Gyr, we repeat  step (3) to again generate hierarchical mergers.
\end{enumerate}

\section{Results}
\label{sect:results}

In this Section, we study repeated mergers, formation of IMBHs, and BHs in the mass gap in a population of NSCs.

\begin{figure} 
\centering
\includegraphics[scale=0.565]{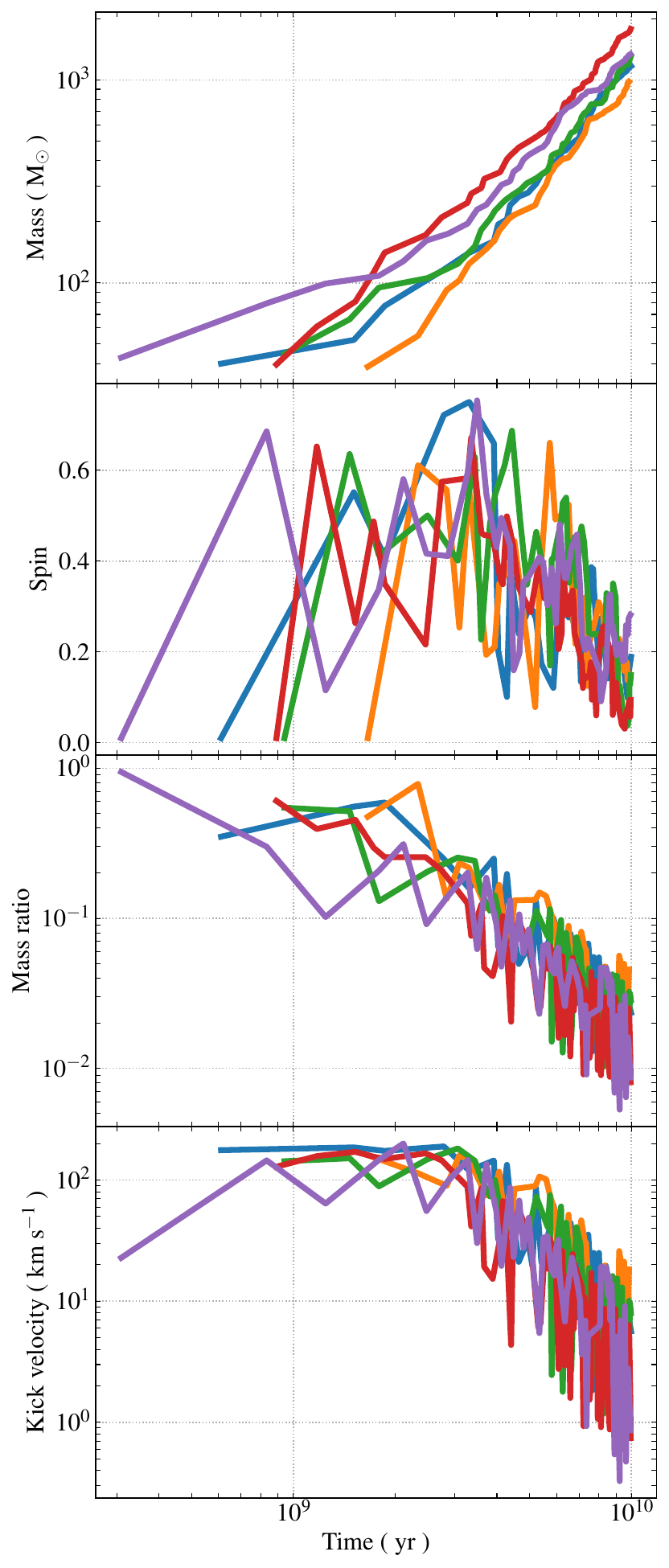}
\caption{Example of growth of $\sim 1000\msun$ IMBHs in Model 1. Different colors represent different formation histories. In order from the top: mass evolution, spin evolution, mass ratio of mergers, recoil kick imparted to the merger remnant. In these examples, the escape speed of the host cluster is around $200\kms$.}
\label{fig:time1}
\end{figure}

We summarize in Table~\ref{tab:models} the models we investigate in our simulations. We explore the role of the host cluster metallicity, primordial binary fraction in massive stars, and BH spin models. We also consider the case the initial seed mass is fixed to four different mass values ($50\msun$, $100\msun$, $150\msun$, $200\msun$). Finally, we run three models where the host cluster mass and density are drawn from distributions that describe a population of GCs. In our simulations, we assume that spin orientations at merger are isotropic, as appropiate to merging binaries assembled in a dynamical environment. We integrate each simulation either up to $10$ Gyr or until the growing seed is ejected from the host cluster, where further growth is quenched. Each result we present is the average over $10^4$ model realizations.

\subsection{Formation of intermediate-mass black holes}

\begin{figure} 
\centering
\includegraphics[scale=0.565]{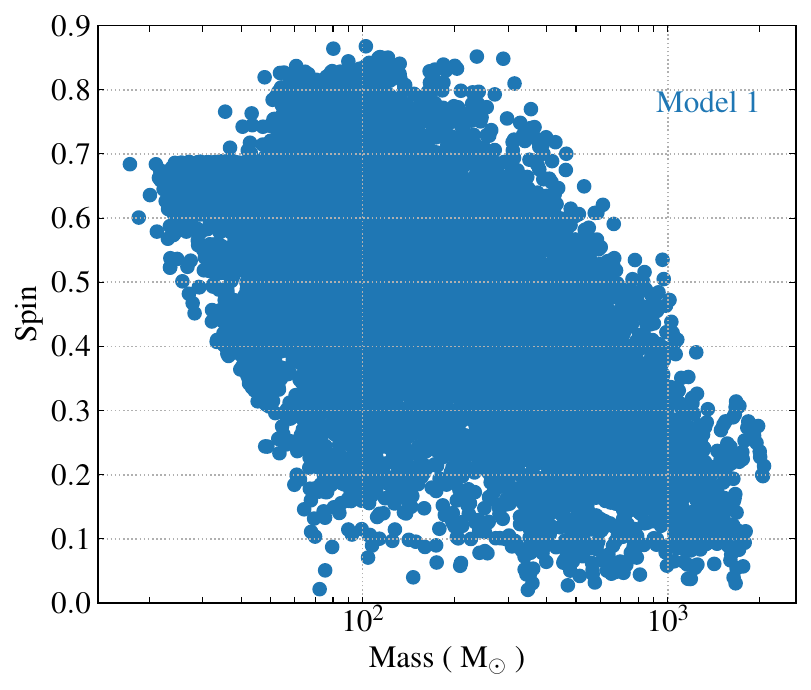}
\caption{Final spin of the growing BH seed as a function of its mass for Model 1. The spin of the BHs tends to decrease with the number of mergers (hence the BH mass) and a correlation exists between mass and spin of the BHs.}
\label{fig:time2}
\end{figure}

We show an example of growth of $\sim 1000\msun$ IMBHs in Model 1, where we plot the mass evolution, the spin evolution, the mass ratio of mergers, and the recoil kick imparted to the merger remnant as a function of time. This is a consequence of the fact that the GW recoil kick imparted to the merger remnant depends both on the asymmetric mass ratio and on the spins of the progenitors (see Eq.~\ref{eqn:vkick}), which we have set to zero in Model 1. As the BH seed merges with other BHs and increases its mass, the typical mass ratio decreases and so does the recoil kick imparted to the growing seed, which remains smaller than the escape speed of the host cluster, which is around $200\kms$ in these examples. 

Concerning the spins, we confirm the previous results that found the spin of the growing seed decreases as a function of time or its growing mass \citep{antonini2019,frsilk2020}. The evolution of the seed spin is quite general, with the first merger producing a remnant with a dimensionless spin parameter of about $0.7$ (starting from two slowly spinning BHs), which then tends to decrease with subsequent mergers. In Figure~\ref{fig:time2}, we show the final BH spin as a function of its mass for Model~1. The spin of the BHs tends to decrease with the number of mergers, eventually producing a negative correlation between mass and spin. The reason is as  follows. After the first mergers, the seed spin is about $0.7$, if BHs are born slowly spinning. The angular momentum imparted to the growing seed during each merger is $\propto m_2\mseed$, being the radius at the innermost stable circular orbit (ISCO) $r_{\rm ISCO}\propto \mseed$, which causes the spin parameter to change by $\sim m_2/\mseed$. Since at early times $m_2$ can be comparable to $\mseed$, the initial variations in the remnant spins tend to be larger than the ones at late times when $\mseed \gg m_2$, as also seen in Figure~\ref{fig:time1}. Assuming an isotropic geometry of BBH mergers as appropriate to a dynamical environment, the final inspiral and deposition of angular momentum happens at random angles to the previous spin axis of the growing seed. However, the growing seed undergoes a damped random walk in the evolution of its spin since retrograde orbits become unstable at a larger specific angular momentum than do prograde orbits, so it is easier to decrease than to increase the spin magnitude \citep{Miller2002,HughesBlandford2003,MandelBrown2008}. So the random walk takes the spin magnitude of the growing seed to small values eventually. Indeed, we find that BHs growing to a mass $\gtrsim 1000\msun$ have dimensionless spin parameters $\lesssim 0.3$.

\begin{figure} 
\centering
\includegraphics[scale=0.565]{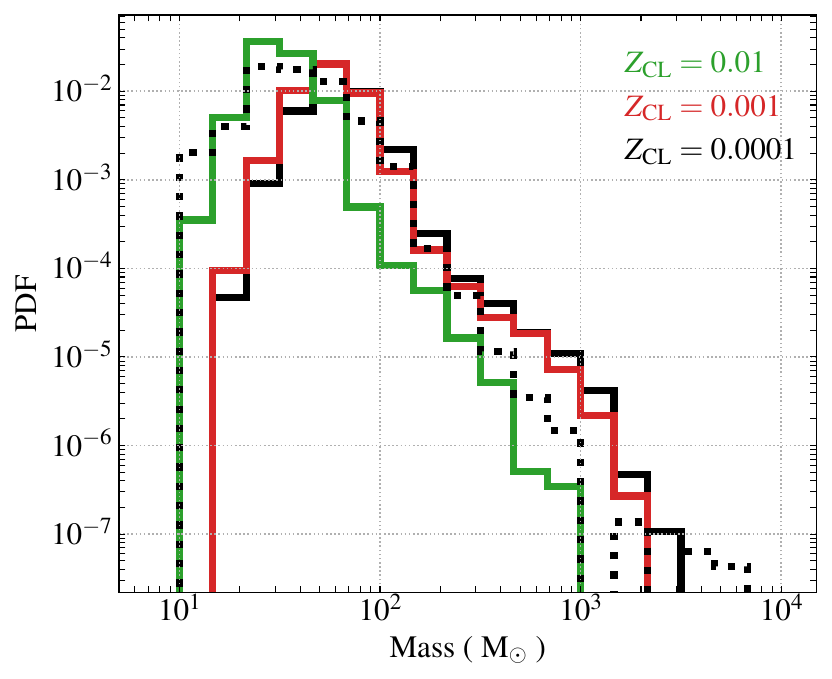}
\includegraphics[scale=0.565]{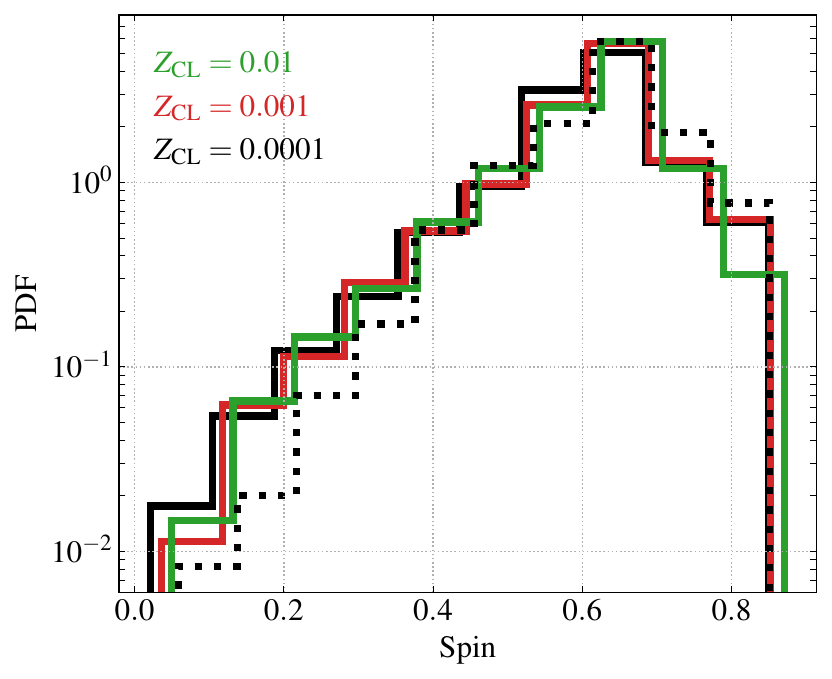}
\caption{Final BH mass (top) and spin (bottom) for different values of cluster metallicity. The dotted-black line represent results from Model 1 when the secondary BH masses are drawn in the range $[m_{\rm min, BH}, \mseed]$. The cluster primordial binary fraction of massive stars is set to $f_{\rm b,*}=0$, while BHs are assumed to be born non-spinning.}
\label{fig:zeta}
\end{figure}

\subsubsection{Effect of the cluster metallicity}

Figure~\ref{fig:zeta} shows the final BH mass (top) and spin (bottom) for different values of cluster metallicity (Models 1-3). In these models, the cluster primordial binary fraction of massive stars to is set to $f_{\rm b,*}=0$, while BHs are assumed to be born non-spinning. The metallicity of the progenitors leaves an imprint on the mass spectrum since it impacts the typical cluster BH mass. We find that the maximum mass that the BH seed can reach is about $1000\msun$ for Solar metallicity and about $3000 \msun$ for $Z_{\rm CL}=0.0001$ and $Z_{\rm CL}=0.001$. Moreover, the peak of the distributions tend to cluster around the maximum mass of the first-generation mergers, which is smaller for higher metallicities. On the other hand, we find that the host cluster metallicity has only a marginal effect on the spin spectrum.

We note that our considerations depend also on the details of the BH mass spectrum. In our models, we have sampled the secondary mass in the range $[m_{\rm min, BH}, m_{\rm max, BH}]$, as appropriate to the BH mass spectrum at a metallicity $Z_{\rm CL}$. To understand how our results depend on this assumption, we re-run Model 1 drawing the secondary masses in the range $[m_{\rm min, BH}, \mseed]$ \citep[as recently assumed in the models of][]{MapelliDall'Amico2021}. This assumption might be justified by the fact that there might be more than one growing seed in a given cluster \citep{KremerSpera2020,GonzalezKremer2021,WeatherfordFragione2021}. While it might be reasonable at the beginning of the cluster evolution, this assumption becomes less justified when the seed mass grows to a few hundreds Solar masses. We illustrate the outcome of this additional run in Figure~\ref{fig:zeta}. We find that the main effect of this choice is to decrease by a factor of $\sim 7$ the relative abundance of seeds that grow to masses $\gtrsim 500\msun$. This trend can be justified by the fact that secondaries can be as massive as $\mseed$ during the whole cluster lifetime, when sampling in the range $[m_{\rm min, BH}, m_{\rm max, BH}]$. As a result, the mass ratio can be as high as unity and the recoil kick can eject the growing BH seed. On the other hand, the ejection of the seed is quenched when sampling in the range $[m_{\rm min, BH}, m_{\rm max, BH}]$, as appropriate to the BH mass spectrum at a metallicity $Z_{\rm CL}$. In this case, the mass ratio becomes too small to lead to high recoil kicks after the seeds have grown to sufficiently high masses \citep{frsilk2020}.

\begin{figure} 
\centering
\includegraphics[scale=0.565]{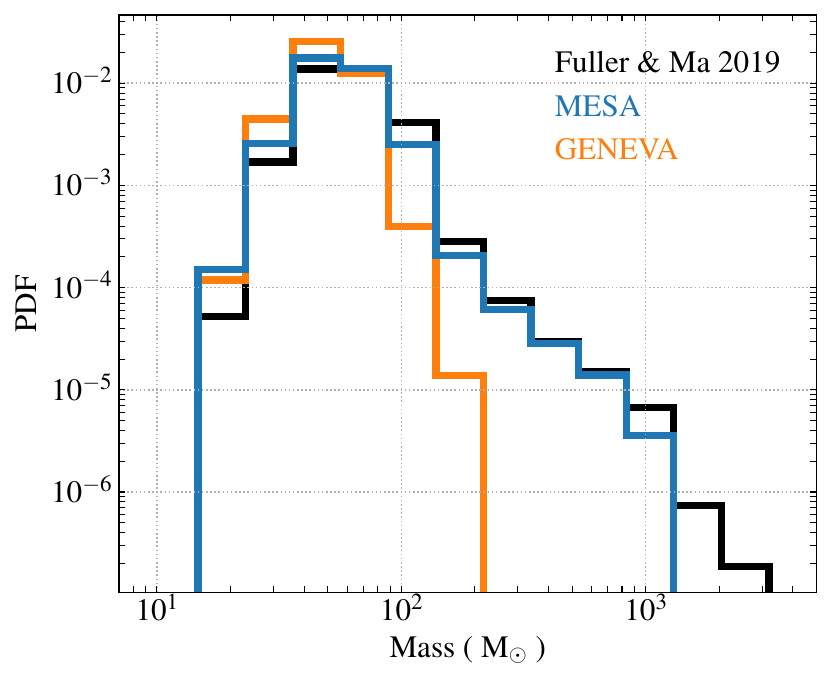}
\caption{Final BH mass for different  prescriptions for BH spins at birth in Models 8-9, compared to Model 1. The cluster metallicity is fixed to $Z_{\rm CL}=0.0001$ and the primordial binary fraction of massive stars is set to $f_{\rm b,*}=0$.}
\label{fig:spin}
\end{figure}

\subsubsection{Effect of primordial binary fraction}

In Models 4--7, we study the effect of the primordial binary fraction for massive stars. In these models, the cluster metallicity is fixed to $Z_{\rm CL}=0.0001$ and BHs are assumed to be born non-spinning, but the primordial binary fraction for massive stars is increased gradually all the way to $f_{\rm b,*}=1$. The main motivation is that bound stellar multiples are common and surveys of massive stars, which are the progenitors of BHs, have shown that more than $\sim 50$\% have at least one or two stellar companions, respectively \citep[e.g,][]{duq91,ragh10,sa2013AA,Sana2017}. We find that changing the primordial binary fractions of stars that are BH progenitors does not have a significant impact on the mass distribution or on the relative outcomes (see Table~\ref{tab:models}).

\subsubsection{Effect of black hole spin}

In Models 8-9, we consider the impact of different assumption on the BH spin at birth. In particular, we use the prescriptions of the \textsc{geneva} stellar evolution code \citep{EggenbergerMeynet2008,EkstromGeorgy2012} and the \textsc{mesa} stellar evolution code \citep{PaxtonBildsten2011,PaxtonMarchant2015} in Model 8 and Model 9, respectively. We show in Figure~\ref{fig:spin} the final BH mass for different  prescriptions for BH spins at birth, compared to Model 1. The cluster metallicity is fixed to $Z_{\rm CL}=0.0001$ and the primordial binary fraction of massive stars to is set to $f_{\rm b,*}=0$. We find that, while Model 9 produces a mass spectrum that is consistent with the mass spectrum from Model 1, Model 8 does not produce BHs more massive than about $200\msun$. The reason is that the \textsc{geneva} code predicts that most of the BHs form with a high spin, while the \textsc{mesa} code leads to BHs of all masses to form with a small residual spin, $\sim 0.1$\footnote{This is because the \textsc{geneva} code do not include magnetic fields, so the core-to-envelope angular momentum transport is purely convective and, therefore, inefficient.}. Therefore, the recoil kicks after merger are typically larger than the cluster escape speed in Model 8, resulting in the ejection of the growing seed. Note also that, while Model 9 produces a mass spectrum that is consistent with the mass spectrum from Model 1, the probability of forming massive BHs is about $1.5$ times smaller than Model 1. The reason has again to be found in the typical BH spins, which are $\sim 0.1$ in Model 9, while BHs are assumed to be born non-spinning in Model 1.

\subsubsection{Initial seed mass}

\begin{figure} 
\centering
\includegraphics[scale=0.565]{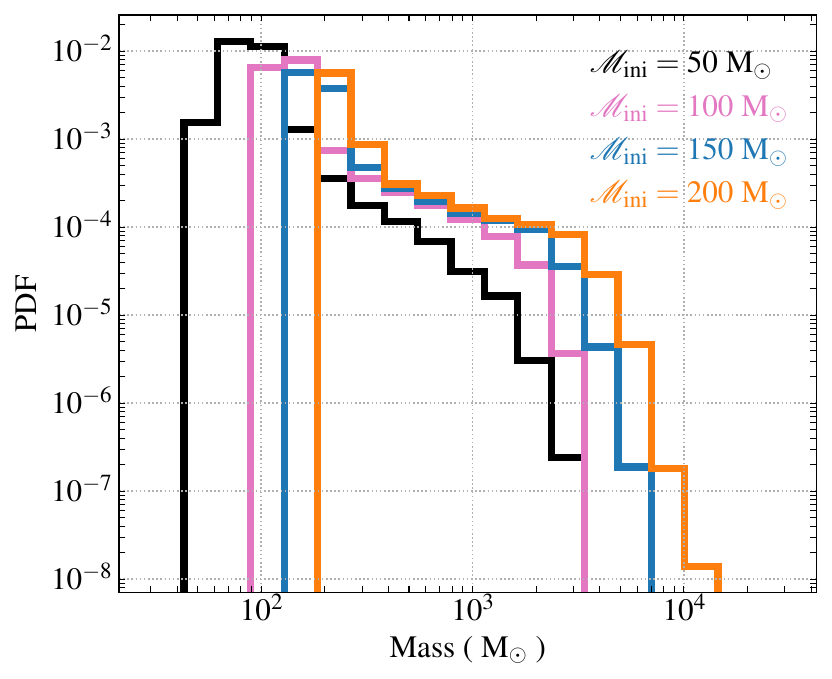}
\caption{Final BH mass for different values of the initial seed mass in Models 10-13. The cluster metallicity is fixed to $Z_{\rm CL}=0.0001$ and the primordial binary fraction of massive stars is set to $f_{\rm b,*}=0$. BHs are assumed to be born non-spinning.}
\label{fig:mseed}
\end{figure}

In Models 10-13, we consider how the final BH mass depends on the initial seed mass $\mseed_{\rm ini}$. We assume that the growing seed has an initial mass of $50\msun$, $100\msun$, $150\msun$, $200\msun$, respectively. In these models, the cluster metallicity is fixed to $Z_{\rm CL}=0.0001$, the primordial binary fraction of massive stars to is set to $f_{\rm b,*}=0$, and BHs are assumed to be born non-spinning. We show the results in Figure~\ref{fig:mseed}. We find that the larger the initial seed mass, the larger the final BH mass. Moreover, about $2\%$, $11\%$, $22\%$, $34\%$ of the runs produce a BH with final mass $>1000\msun$ for $\mseed_{\rm ini}=50\msun$, $100\msun$, $150\msun$, $200\msun$, respectively (see Table~\ref{tab:models}). Therefore, if clusters are born with a massive seed \citep[e.g.,][]{KremerSpera2020,GonzalezKremer2021,WeatherfordFragione2021}, it can grow efficiently to $\gtrsim 10^3\msun$.

\subsubsection{Cluster evolution}

\begin{figure} 
\centering
\includegraphics[scale=0.565]{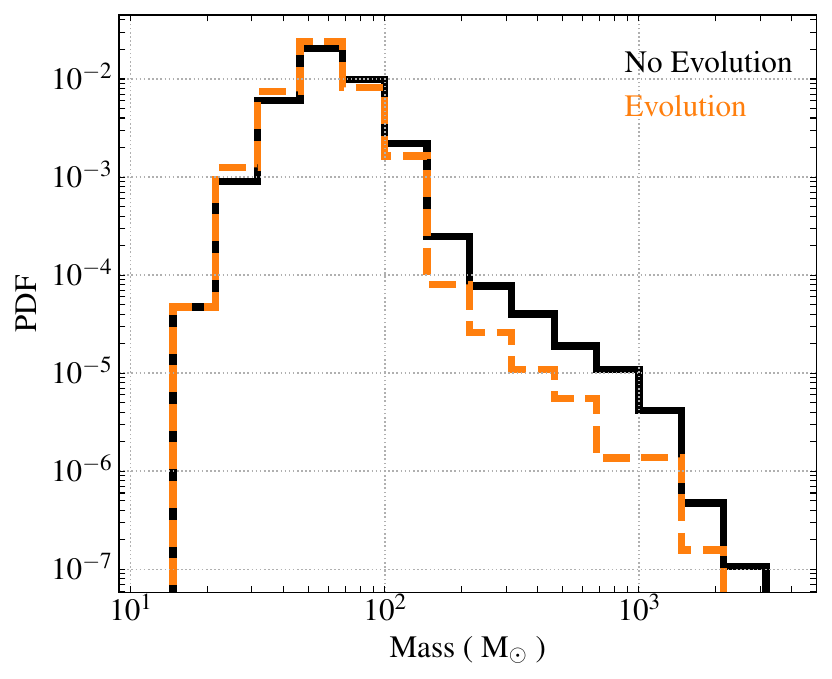}
\caption{Final BH mass in Models 1 without (black solid) and with (orange dashed) cluster evolution. The cluster metallicity is fixed to $Z_{\rm CL}=0.01$ and the primordial binary fraction of massive stars is set to $f_{\rm b,*}=0$, while BHs are assumed to be born non-spinning.}
\label{fig:evol}
\end{figure}

In our approach, we have neglected the possible evolution of the global properties of the NSCs. While there is not a straightforward way to model the cosmic evolution of NSCs, we can adopt a simplified model (Model 1E in Table~\ref{tab:models}) similar to what is sometimes done for GCs, which experience a less complex evolution. Specifically, we follow  H\'{e}non's principle to model the global properties of a cluster and assume that the heating rate from BH binaries in the core has to balance the energy flow into the whole cluster \citep{Henon1961}. We model the heating rate of the star cluster as a function of time as \citep{GielesHeggie2011,AlexanderGieles2012,antonini2019,AntoniniGieles2020}.
\begin{equation}
\dot{E}(t)=\dot{E}(0)\left(\frac{3}{2} \frac{\gamma t}{\tau_{\rm rh,0}} + 1\right)^{-5/3}\,,
\label{eqn:evvv}
\end{equation}
where $\gamma \approx 0.1$ is a constant,
\begin{eqnarray}
\dot{E}(0) &=& 2.3 \times 10^5 \msun\,(\kms)^{2}\,{\rm Myr}^{-1}\nonumber\\
&\times & \left(\frac{M_{\rm CL}}{10^5\msun}\right)^{2/3}\left(\frac{\rho_{\rm h}}{10^5\msun\ \mathrm{pc}^{-3}}\right)^{5/6}\,,
\end{eqnarray}
$\rho_{\rm h}=3M_{\rm CL}/(8\pi r_{\rm h}^3)$ is the average density within the cluster half-mass radius, and $\tau_{\rm rh,0}$ is the initial average relaxation timescale within $r_{\rm h}$
\begin{equation}
\tau_{\rm rh,0}=7.5\,{\rm Myr}\left(\frac{M_{\rm CL}}{10^5\msun}\right)\left(\frac{10^5\msun\ \mathrm{pc}^{-3}}{\rho_{\rm h}}\right)^{1/2}\,.
\end{equation}
H\'{e}non's principle imposes that the required heating rate of the cluster is balanced with the loss of energy from the BH binaries in the core, and we assume that the binary containing the growing seed dominates the heating at all times. Therefore, we require that $\dot{E}_{\rm bin}(t)=-\dot{E}(t)$, where $\dot{E}_{\rm bin}(t)$ is the rate of energy loss from the binary. From the previous equation, it follows that the timescale during which the binary exists and dynamical interactions dominate the energy flow of the cluster is given by \citep[][]{antonini2019}
\begin{equation}
\tau_{\rm bin} = \frac{G\mseed m_2}{2\max(a_{\rm ej},a_{\rm GW}) \dot{E}(t)}\,.
\label{eqn:taubin}
\end{equation}
Again, $\max(a_{\rm ej},a_{\rm GW})$ (see Eqs.~\ref{eqn:aej}-~\ref{eqn:agw}) is the semi-major axis at which the hardening interactions stop as a result of either a merger or the ejection of the binary. Now, $a_{\rm GW}$ is computed by requiring that the rate of energy loss due to dynamical hardening equals that due to GW radiation\footnote{Note that we are neglecting a factor $(1-e_{\rm BH})^{-7/10}$, which is of order unity.} \citep{antonini2019}
\begin{equation}
a_{\rm GW}=0.01\left[\frac{(\mseed m_2)^2(\mseed+m_2)}{M_\odot^5}\frac{\msun\,(\kms)^{2}\,{\rm Myr}^{-1}}{\dot{E}_{\rm bin}(t)}\right]^{1/5}    \,.
\end{equation}
From Eq.~\ref{eqn:taubin}, we estimate the timescale for the binary to shrink to $\max(a_{\rm ej},a_{\rm GW})$; if $a_{\rm ej}>a_{\rm GW}$, the binary is ejected from the cluster halting further growth, otherwise the merger happens in cluster. Note that in this framework the escape velocity decreases as a function of time as the cluster evolves
\begin{equation}
\vesc(t)=v_{\rm esc, 0}\left(\frac{3}{2} \frac{\gamma t}{\tau_{\rm rh,0}} + 1\right)^{-1/3}\,,
\end{equation}
where $v_{\rm esc, 0}$ is the initial escape velocity. 

Figure~\ref{fig:evol} compares the final BH mass obtained in Model 1 to the same model when a model (Eq.~\ref{eqn:evvv}) for cluster evolution is taken into account. We find that including cluster evolution does not change significantly our results for masses $\lesssim 100\msun$. On the other hand, the likelihood of producing a BH with final mass $\gtrsim 100\msun$ decreases by a factor of a few (see Table~\ref{tab:models}), owing to a typically longer binary heating rate (with respect to the binary shrinking rate computed with no cluster evolution) and decreasing cluster escape velocity.

\subsection{Black holes in mass gap}

\begin{figure} 
\centering
\includegraphics[scale=0.565]{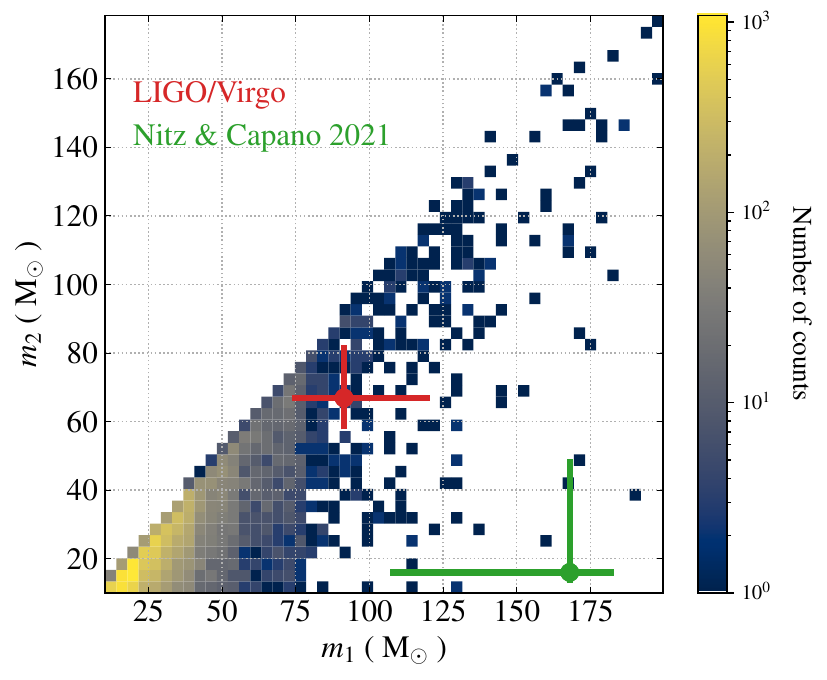}
\includegraphics[scale=0.565]{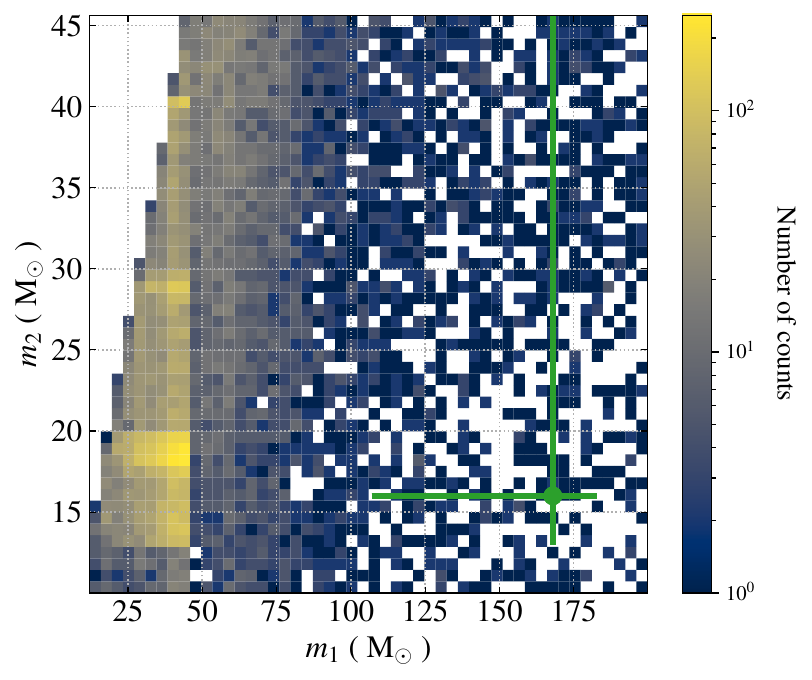}
\caption{Mass of secondary versus primary component of BBH mergers in Model 1, when the secondary BH masses are drawn in the range $[m_{\rm min, BH}, \mseed]$ (top) and in the range $[m_{\rm min, BH}, m_{\rm max, BH}]$ (bottom). We also report the estimated component masses of GW190521 from LVK collaboration \citep{lvc2020cat} and from \citet{NitzCapano2021}.}
\label{fig:gw190521}
\end{figure}

Here we show how our semi-analytical framework predicts black hole mergers in the mass gap. As an example, we explore the possibility that a system like GW190521 formed through multiple mergers. We show in Figure~\ref{fig:gw190521} (top panel) the mass the secondary versus primary component of binary BH mergers in Model~1, when the secondary BH masses are drawn in the range $[m_{\rm min, BH}, \mseed]$. This assumption is important since the maximum BH mass, even at low metallicity, is around $50\msun$ in our models, which requires the secondary of GW190521 to be the product of a previous merger as well \citep[e.g.,][]{FragioneLoeb2020,KimballTalbot2020,KimballTalbot2020b,MapelliDall'Amico2021}. We also show the estimated component masses of GW190521 \citep{lvc2020cat}. Our results
confirm that repeated mergers in dense star clusters are able to match the component masses of GW190521. Recently, \citet{NitzCapano2021} reanalyzed the data of the LVK collaboration with a
new waveform allowing for more extreme mass ratios and found that GW190521 is consistent with an intermediate-mass ratio inspiral with primary mass $168^{+15}_{-61}\msun$ and secondary mass $16^{+33}_{-3}\msun$ \citep[see also][]{MehtaBuonanno2021}. In this case, the primary mass is not inside the mass gap, but it is in the IMBH regime, while the secondary mass could be an ordinary (first-generation) stellar BH. As we show in the bottom panel of Figure~\ref{fig:gw190521}, GW190521 would then be more consistent with Model~1 when the secondary BH masses are drawn in the range $[m_{\rm min, BH}, m_{\rm max, BH}]$, as appropriate to the cluster metallicity $Z_{\rm CL}$.

\begin{figure} 
\centering
\includegraphics[scale=0.565]{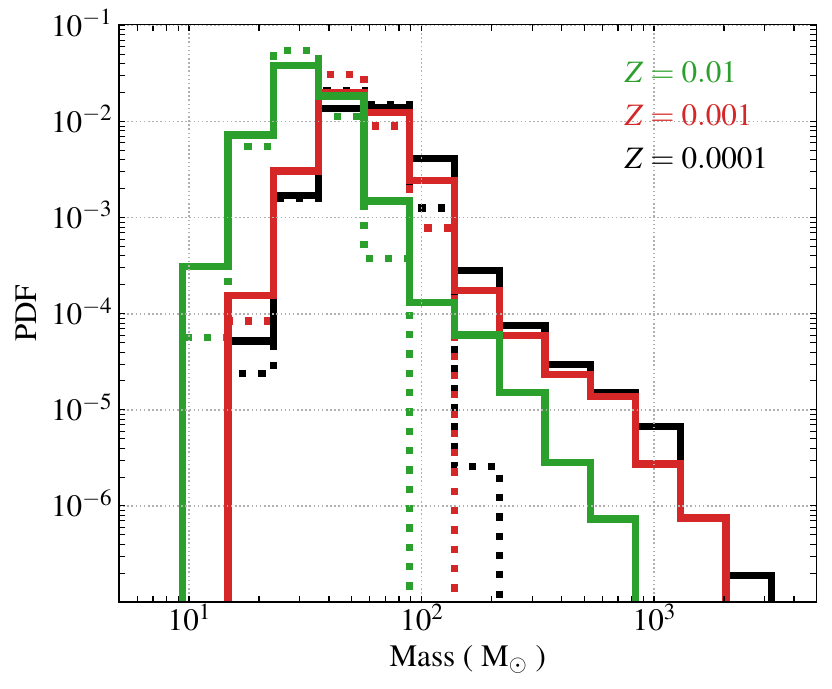}
\caption{Final BH mass for different values of the cluster metallicity for NSCs (solid) from Models~1--3 and GCs (dotted) from Models~14--16. The primordial binary fraction in massive stars is set to $f_{\rm b,*}=0$, while BHs are assumed to be born non-spinning.}
\label{fig:gc}
\end{figure}

\subsection{Globular clusters}

For comparison, Models 14--16 represent GCs as host star clusters. For these models we draw the total masses from a log-normal distribution with mean $\langle \log_{10} (M_{\rm CL}/\msun) \rangle = 5.3$
and standard deviation $0.4$ \citep{Harris1996}. Following \citet{gne14}, we adopt the average density at the half-mass radius
\begin{equation}
\rho_{\rm h}=
\begin{cases}
10^3\msun\, \mathrm{pc}^{-3}& {\rm for\;} M \le 10^5\msun \cr
10^3 \left(\frac{M}{10^5\msun}\right)^2\msun\, \mathrm{pc}^{-3}& {\rm for\;} 10^5\msun <  M < 10^6\msun \cr
10^5\msun\, \mathrm{pc}^{-3}& {\rm for\;} M \ge 10^6\msun\,.
\end{cases}
\end{equation}
The above equation limits $\rho_{\rm h}$ to $10^5\msun\,\mathrm{pc}^{-3}$ in the most massive clusters, that is about the highest observed half-mass density. The cluster primordial binary fraction for massive stars is set to $f_{\rm b,*}=0$, while BHs are assumed to be born non-spinning.

We show the final BH mass for different values of the cluster metallicity for GCs in Figure~\ref{fig:gc} and compare to the results from Models~1--3. We find that only about $3\%$ and $2\%$ of the runs produce a BH more massive than $100\msun$ for $Z_{\rm CL}=0.0001$ and $0.001$, while none in the case of solar metallicity (see Table~\ref{tab:models}). The reason is of course that GCs have lower escape speeds compared to NSCs and the recoil kick can eject the growing seed easily, halting further growth.

\section{Discussion and conclusions}
\label{sect:concl}

Our methodology is based on some approximations and assumptions. We discuss them in the following section and conclude with a brief summary of our main results.

\subsection{Caveats and improvements in current methodology}

In our models, we have sampled the masses and half-mass radii of NSCs from Eqs.~\ref{eqn:mclmgal}-\ref{eqn:rhmcl}, which are calibrated over observations of NSCs in the local Universe \citep{georg2016}. We have assumed that these properties have not evolved during the lifetime of a given NSC. However, star clusters expand as a consequence of two-body relaxation and lose mass (and expand) as a result of dynamical ejections and stellar evolution. For example, the fraction of mass in stars that turn into BHs is about $11\%$, for a canonical \citep{kro01} IMF. For low metallicities, about $2/3$ of the progenitor mass will be lost during stellar evolution, while this fraction is about $9/10$, for solar metallicity \citep[e.g., see Figure~1 in][]{FragioneLoeb2020}. Therefore, a given cluster will lose about $7\%$ and $10\%$ of its initial mass for low and solar metallicities, respectively, when BH progenitors collapse and form BHs. This mass loss (along with the mass lost by other stars) will cause the star cluster to expand during its lifetime. Cluster expansion tend to lower the cluster mass and density, which in turn could prevent the growth of a massive seed \citep[e.g.,][]{antonini2019}. However, episodes of core collapse and of gravothermal oscillations could temporary increase the central density, which could result in an enhancement of the growth of the BH seed \citep[e.g.,][]{BreenHeggie2013}. Therefore, the overall effect of including the evolution of the NSC is difficult to predict.

While the previous discussion essentially assumes that a NSC is isolated, other important factors could be relevant to place a population of NSCs in the proper galactic and cosmological context. For example, we do not model the evolution of NSCs using an $N$-body approach, as this  is computationally expensive and beyond what current codes can handle \citep[e.g.,][]{Aarseth2003,gier2006MNRAS.371..484G,RodriguezWeatherford2021}. While there is not a straightforward way to model the cosmic evolution of NSCs, various simplified models could be adopted \citep{GielesHeggie2011,antonini2019}. However, NSCs are not isolated environments on a cosmological scale, as there can be continuous supply of stars and gas from the rest of the galaxy \citep[e.g.,][]{VogelsbergerGenel2014}, there could be ongoing star formation and accretion of smaller star clusters \citep[e.g.,][]{ant2014}, and SMBHs could be delivered to this region following galaxy collisions \citep[e.g.,][]{BegelmanBlandford1980}. Note also that the NSC evolution could be significantly affected by the presence of a (single or binary) SMBH, expecially if SMBH seeds are formed via direct collapse in the early Universe. For example, SMBHs tend to create a cusp of stars and compact objects \citep[e.g.,][]{bahc1976,hopale2006,alex2017,frasar2018} and could accrete gas \citep[e.g.][]{kroupa2020,Natarajan2021}, with an AGN component that could be complementary in some cases \citep{TagawaKocsis2021a}.

Finally, we note that we have not included star-star collisions and stellar runaway mergers. These processes could play a role since could form a massive BH seed from the collapse of a very massive star and, eventually, further accretion \citep[e.g.,][]{por02,InayoshiVisbal2020,KremerSpera2020,TagawaHaiman2020,DiCarloMapelli2021}. These processes could possibly be enhanced in the case of a large primordial binary fraction of massive stars \citep{GonzalezKremer2021} and in the case of a top-heavy initial mass function \citep{WeatherfordFragione2021}. Tidal captures of stars may also lead to the formation of a massive BH \citep{2017MNRAS.467.4180S}. For the typical masses and sizes of the NSCs in our sample, the timescales for this process and repeated mergers could be comparable. However, the tidal capture process relies on the assumption that a few BHs lurk in the core, which might not be the case since the BH-burning process, which ejects most of the cluster BHs, could last more than a Hubble time for a typical NSC \citep[e.g.,][]{KremerYe2020}.

Despite all the above limitations, our model gives general scenarios for the formation of massive BHs through repeated mergers in dense star clusters, and the population of ejected BHs \citep[see also][]{frsilk2020}. Importantly, we have drawn masses and half-mass radii of NSCs consistently with observations \citep{georg2016}, weighting their abundances using the distribution of galaxies in the local Universe \citep{FurlongBower2015}. This procedure is crucial to ensure the correct sampling and correlation of the distributions of NSC parameters, since their masses and sizes are closely related. While our analysis still lacks a detailed prescription for the evolution of the background cluster, we have included some treatment of our dense star clusters as a multi-zone model, using the three-parameter potential-density pair of \citet{StoneOstriker2015}.

We have fully integrated for the first time a (single and binary) stellar evolution code with the underlying dynamical model. We note that our current stellar evolutionary code includes the most up-to-date prescriptions for stellar winds and remnant formation \citep{BanerjeeBelczynski2020}. This allows us to compute a realistic BH mass spectrum for any cluster metallicity and to determine BH spins at birth from fits to stellar tracks \citep{EggenbergerMeynet2008,PaxtonBildsten2011,EkstromGeorgy2012,PaxtonMarchant2015}. Both the distributions of initial BH masses and spins are fundamental to determine if a massive seed can grow as a result of hierarchical mergers. For the sake of versatility, our method also includes the possibility of including an initial BH seed (with a given spin), which might be the end product of a very massive star formed via stellar mergers \citep[e.g.,][]{2002ApJ...576..899P,KremerSpera2020,DiCarloMapelli2021,GonzalezKremer2021}.

Finally, we note that our scheme still lacks the time evolution of the BH mass spectrum, which can be modified over time by dynamical ejection of BHs. Also, we take into account only mergers between our growing seed and first-generation BHs, neglecting the possible contributions of secondaries from a previous merger. Finally, we assume that our processes take place after an average timescale, which could be in principle replaced by a direct integration of the few-body interaction continuously experienced by the growing seed. We leave the detailed exploration and implementation of each of these improvements to future work.

\subsection{Summary}

The mass spectrum of BHs is among the most hotly debated topics in modern astrophysics. Current stellar evolution models predict a dearth of BHs  with masses $\gtrsim 50\msun$ and, even though the LVK collaboration has detected almost fifty BBHs, the exact shape of the BH mass spectrum remains a mystery.

In this paper, we have presented a semi-analytic framework to investigate hierarchical mergers in dense star clusters, expanding upon the method originally developed in \citet{frsilk2020}. Our method allows us to rapidly study the outcomes of hierarchical mergers in dense star clusters and probe how they are effected by the cluster masses, densities, metallicities, and primordial binary fractions and by the different assumptions on the BH mass spectrum and spins.

We have studied repeated mergers, BHs in the mass gap, and formation of IMBHs in a population of NSCs. We have shown that a BH seed can grow up to $\sim 10^3\msun$-$10^4\msun$ as a result of repeated mergers with stellar BHs. We have found that the smaller the cluster metallicity, the larger is the typical final BH mass. Moreover, we have confirmed that the spin of the BHs tends to decrease with the number of mergers and, as a result, more massive BHs have typically smaller spins. We have also found that the primordial binary fraction in massive stars does not have a significant impact on our results, while the choice of the BH spin model is crucial, with low birth spins leading to the production of a larger population of massive BHs. We have also illustrated that GW190521 can be born as a result of hierarchical mergers in a NSC. Finally, we have discussed that, unlike NSCs, GCs are not massive and dense enough to retain the merger remnant after a recoil kick is imparted, and a BH seed cannot grow significantly in mass via repeated mergers with other BHs.

\section*{Acknowledgements}

We thank Sambaran Banerjee for useful discussions on stellar evolutionary codes and for updating \textsc{sse} and \textsc{bse} with state-of-the-art prescriptions. G.F.\ and F.A.R.\ acknowledge support from NASA Grant 80NSSC21K1722. This work received funding from the European Research Council (ERC) under the European Union’s Horizon 2020 Programme for Research and Innovation ERC-2014-STG under grant agreement No. 638435 (GalNUC) (to B.K.).

\bibliography{refs}

\end{document}